# Document Clustering based on Topic Maps

Muhammad Rafi
Assistant Professor

National University of Computer & Emerging Sciences,

Karachi Campus, Pakistan

M. Shahid Shaikh
Associate Professor

National University of Computer & Emerging Sciences,

Karachi Campus, Pakistan

Amir Farooq
National University of Computer & Emerging Sciences,

Karachi Campus, Pakistan

## ABSTRACT
Importance of document clustering is now widely acknowledged by researchers for better management, smart navigation, efficient filtering, and concise summarization of large collection of documents like World Wide Web (WWW). The next challenge lies in semantically performing clustering based on the semantic contents of the document. The problem of document clustering has two main components: (1) to represent the document in such a form that inherently captures semantics of the text. This may also help to reduce dimensionality of the document, and (2) to define a similarity measure based on the semantic representation such that it assigns higher numerical values to document pairs which have higher semantic relationship. Feature space of the documents can be very challenging for document clustering. A document may contain multiple topics, it may contain a large set of class-independent general-words, and a handful class-specific core-words. With these features in mind, traditional agglomerative clustering algorithms, which are based on either Document Vector model (DVM) or Suffix Tree model (STC), are less efficient in producing results with high cluster quality. This paper introduces a new approach for document clustering based on the Topic Map representation of the documents. The document is being transformed into a compact form. A similarity measure is proposed based upon the inferred information through topic maps data and structures. The suggested method is implemented using agglomerative hierarchal clustering and tested on standard Information retrieval (IR) datasets. The comparative experiment reveals that the proposed approach is effective in improving the cluster quality.

## 1. INTRODUCTION
Clustering as an unsupervised data mining approach is widely used in variety of situations. It automatically groups a collection into meaningful sub-groups; the word meaningful is rather relative. The challenging part is to extract the meaningfulness and to control the objective of the "best" clustering sense. Document clustering, is a specialized data clustering problem, where the objects are in the form of documents. The objective of the clustering process is to group the documents which are similar in some sense like: type of document, contents of document, etc into a single group (cluster). The difficult part is to learn from a data set, actually how many classes of such groups exist in the collection.

Document Clustering aims to discover natural grouping among documents in such a way that documents with in a cluster are similar (high intra cluster similarity) to one another and are dissimilar to documents in other clusters (low inter cluster similarity). Exploring, analyzing and correctly classifying the unknown natures of data in a document without supervision is the major requirement of document clustering method. Traditionally, document clustering algorithms mainly uses features like: words, phrases, and sequences from the documents to perform cluster. These algorithms generally apply simple features extraction techniques that mainly based on feature counting and frequency distribution of the features to decide about the relatedness among documents. All these approaches thus, do not able to cater the meaning behind the text (words). These techniques simply perform clustering independent of the context. Document written in human language contains a context and the usage of words are largely depends on the context of the written text. Recently, few researchers have suggested some different model for document representation that captures the inherit semantics of the words. Like Phrase-based and common sequence of word based approaches. These have reported outstanding results in document clustering. In this paper, we introduced a new document representation model based on the information topic maps that are present in a document. Topic Maps is becoming an international standard for knowledge representation that facilitates the find ability of information later. It is based on formal model and modern information management. The model is very well defined in ISO standard (ISO13250). The use of topic maps is escalating in the projects that perform enterprise information integration, knowledge management, digital libraries and web-based information integration and management. To the best of our knowledge this is very first attempt that uses these topic maps data structures in document clustering. The subject-centric nature of topic-maps paradigm is what we perceive the very nature of human central clustering process. We wish to exploits the very nature of this subject centric information in topic maps for performing the final clustering. The previous attempts like: vector based document clustering uses the document centric nature for clustering, similarly frequent pattern based or frequent common sequence based approaches also uses document centric and application centric view for performing clustering. Topic maps offers an out of the box meta- model for enabling subject based information management. The algorithm suggested in this paper is specially suited for multi-topic documents. The key feature of this approach to represent the document in a compact topic map based data structures, which ultimately reduce the size of the document in an effective manner. Further, the topic extracted data sets are used to define the similarity measure for a pair of documents. Finally, the Hierarchical agglomerative clustering is used to perform the clustering. The results of the standard clustering measures produced in this study using the standard information retrieval datasets, clearly outperform the previously known approaches to semantics based document clustering. In the section three, we discuss the related works of this study; next we discuss our approach to document clustering. Then we discuss the experimental setup, data set and the measures of this study. Finally, in the last section we discuss the results and conclusion of the work.

## 2. LITERATURE REVIEW

Data clustering [1] is an unsupervised technique for discovering valuable knowledge from data. Document clustering, is a specialized data clustering problem, where the objects are in the form of documents. The objective of the clustering process is to group the documents which are similar in some sense like: type of document, contents of document, etc into a single group (cluster). The difficult part is to learn from a data set, actually how many classes of such groups exist in the collection. Document Clustering aims to discover natural grouping among documents in such a way that documents with in a cluster are similar (high intra cluster similarity) to one another and are dissimilar to documents in other clusters (low inter cluster similarity). Exploring, analyzing and correctly classifying the unknown natures of data in a document without supervision is the major requirement of document clustering method.

Clustering is an effective method for search computing [2]. It offers the possibilities like: grouping similar results [3], comprehend the links between the results [4] and creating the succinct representation and display of search results.

Traditionally, document clustering algorithms mainly uses features like: words, phrases, and sequences from the documents to perform clustering [5,6,7,8]. These algorithms generally apply simple features extraction techniques that mainly based on feature counting and frequency distribution of the features to decide about the relatedness among documents. All these approaches thus, do not able to cater the meaning behind the text (words). These techniques simply perform clustering independent of the context. Document written in human language contains a context and the usage of words are largely depends on this context.

Topic Maps [9] becoming a standard for describing knowledge structures and using it to support later the find ability of that coded knowledge. The main emphasis of topic maps structures is to develop a vis-à-vis relation among the knowledge contents. The original thrust for the creation of topic maps is to merge back-of-book indexes. It is a vigorous tool for merging information from both structured and unstructured form whether it is in the form of glossaries, cross-references, thesauri, or catalogs. A topic map is a representation of a set of assertions about one or more subjects. There are in fact three kinds of assertions topic names, occurrences and associations. The information structure of Topic Maps called TAO model [10] is used to structured information in topic maps format. The major benefits of using topic maps for document clustering task are (1) it can reduce the size of document (2) it can capture the topic related information from the document in an structured form (3) the inherit nature of arbitrary and robust information merging and (4) it can easily handle the semantic topics and its hidden relationship and associations. In our proposed method, we have used the topic maps for the representation of documents, we also proposed a similarity measure based on this representation and finally for the sake of clustering we used hierarchical agglomerative clustering (HAC) algorithm. The performance comparison of clustering algorithms can be done on various parameters. The comparison perform in [12,13] are the standard measures for evaluating an algorithm on document clustering. We have used the same parameters for evaluating study on the proposed method.

## 3. DOCUMENT CLUSTERING BASED ON TOPIC MAPS (TMHC)

There are three basic steps involved in document clustering; our algorithm is also following these steps. We first transform each document in a compact form which only represents the topics presented in the document along with the occurrence and association between topics. The topic maps information is generated by using an online open source application Wandora [14]. It is an application for applying data analysis techniques on topic maps and is able to generate topic maps supporting various data representations. It uses an online plug-in, integrated with a service Open Calais which proved to be very useful in generating Topic maps, taking plain text files as input and returned topic maps based on the information present in these text files. The topic maps where then exported into XTM format using the Wandora's export Utility. After collection of topic maps, a similarity measure was developed which extracted useful information from the topic maps, the three levels of information was identified in the topic maps to be used as the criterion for clustering. First the major topics which were assigned to a document, secondly the tags represent informational terms such as Country, City, Technology, Designations, etc and after that the actual values corresponding to these tags such as Pakistan, Karachi etc. Xpath Queries were used to extract relevant topics, tags and their values from the XTM Files. A document to document similarity matrix is developed using this measure. Finally the hierarchical agglomerative clustering is applied to obtain the final clusters. Figure 1: shows the steps involved in topic maps based document clustering approach (TMHC).

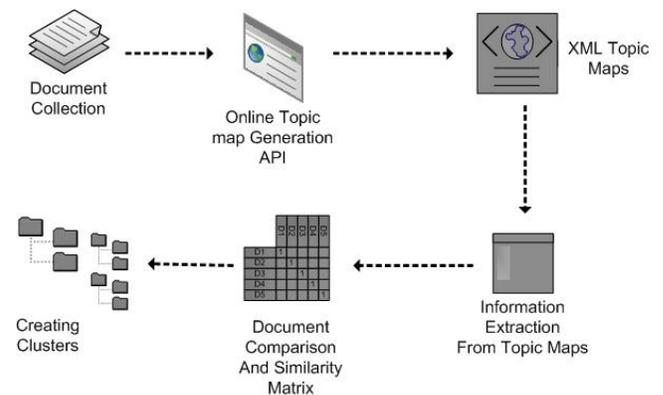

**Figure 1: Steps involved in TMHC**

## 3.1 Document Representation in TM

The three datasets contains different formats for documents, all the three datasets are parsed to transform in topic maps based data structures by using Wandora. The topics and their related information constructs are then extracted from XML Topic Maps representation by using XPath expressions.

## 3.2 Similarity Measure based on TM

A similarity measure is a numerical value in between 0 and 1 that represent the relatedness between a pair of documents. The document transformed into TM representation, carries three valuable constructs: 1. Topic 2. Topic-Tag and 3. Tag value. The similarity we defined is being calculated as the number of common topics, with common topic tags and the tag-value. The following equation calculates the similarity between a pair of document $D_i$ and $D_j$. where $DT_i$ and $DT_j$ represent the common topic presented in the document that is successfully extracted from the topic map representation. The topic-tag is the instance of a topic for example; sports is a topic, cricket, and hockey are topic tags. Hence *Dtagi* and *Dtagj* are the common topic tags. The last construct that must be incorporated in similarity is the tag-value. It is the instance level data that is presented for any topic. So, *DtagVali* and *DtagValj* respectively represent these values.

$$sim(Di, Dj) = (\sum(DTi, DTj) + \sum(Dtagi, Dtagj) + \sum(DtagVali, DtagValj )) / (totalDT + totalDtag + totalDtagVal )$$

The above equation has been used to compute similarity score among pair of documents.

## 4. EXPERIMENTAL SETUP

The proposed algorithmic approach has been compared with a bunch of recently proposed document clustering algorithms on some of the standard dataset for the problem of document clustering. We have implemented the FIHC and CFWS in our previous work [13]. The TMHC is implemented in Java programming language. The experiment is executed on a Dell Vostro 1520 Notebook, with Intel Core2duo processor and 2GB of RAM with 200GB of Hard Disk storage.

## 4.1 Datasets

The three standard document datasets are used to compare the quality of our clustering method. These three data sets are selected mainly due to the fact that most researchers whom work is related to this study have used the same datasets to report their results and comparisons. These datasets are:

▪ Reuters: The Reuters-21578, test collection of Distribution 1.0 is used. The collection appeared in Reuter's newswire in the year 1987. The collection consists of 22 data files, an SGML DTD file describing the format of the available data, and six files describing the categories used to index data. The collection is available at http://www.daviddlewis.com/resources/testcollections/reuters 21578/

▪ NEWS20: It is also a popular data set among text mining community; it's mainly used for text classification and clustering measure for machine learning techniques. The data set consists of approximately 20,000 newsgroup documents, partitioned in 20 different classes. The data set is available at http://people.csail.mit.edu/jrennie/20Newsgroups/

▪ OHSUMED: The OHSUMED collection of the 1987-1991 abstract from 270 journals. It consists of over 348,566 references from the MEDLINE database, which is a database of medical literature maintained by the National Library of Medicine (NLM). Most of the references have abstracts and all have associated MeSH (Medical Subject Headings) indexing terms, with some of the MeSH terms marked as primary. The data set is available at http://davis.wpi.edu/xmdv/datasets/ohsumed.html

We have selected subsets from these data sets for measuring the quality of results produced by our algorithm.

**Table 1: Experimental Datasets**

| Data Sets | Data Source | No. of Docs | No. of Classes |
|-----------|-------------|-------------|----------------|
| **D1** | Reuters 21578 | 4650 | 52 |
| **D2** | Reuters 21578 | 1661 | 23 |
| **D3** | Reuters 21578 | 797 | 15 |
| **D4** | NEWS20 | 1203 | 07 |
| **D5** | OHSUMED | 1240 | 11 |
| **D6** | OHSUMED | 756 | 09 |

## 4.2 Evaluation Measures

We justify the effectiveness of our proposed method by using standard cluster quality measures like F-Measure, Purity and Entropy.

### 4.2.1 F-Measure

The F-measure uses a combination of precision and recall values of clusters. We let $n_i$ designate the number of documents in class $i$, and $c_j$ designate the number of documents in cluster $j$. Moreover, we let $c_{ij}$ designate the number of items of class $i$ present in cluster $j$. Then we can define *prec(i, j)*, the precision of cluster $j$ with respect to class $i$ and *rec(i,j)*, the recall of a cluster $j$ with respect to class $i$ as $prec(i,j) = \frac{c_{ij}}{c_j}$ and $rec(i,j) = \frac{c_{ij}}{n_i}$. The F-measure, *F(i,j)*, of a class $i$ with respect to cluster $j$ is then defined as

$$F(i,j) = \frac{2 * prec(i,j) * rec(i,j)}{prec(i,j) + rec(i,j)}$$

The f-measure for the entire clustering result is defined as

$$F = \sum_i \frac{n_i}{n} \max(F(i,j))$$

### 4.2.2 Purity

Purity can be defined as the maximal precision value for each class j, We compute the purity for a cluster $j$ as $purity(j) = \frac{1}{c_j} max(c_{ij})$. We then define the purity of the entire clustering result as:

$$Purity = \sum_j \frac{c_j}{N} purity(j)$$

Where $N = \sum_j c_j$, i.e. the sum of the cardinalities of each cluster, Note that we use this quantity rather than the size of the document collection for computing the purity.

### 4.2.3 Entropy

Entropy measure how homogenous each cluster j is. It can be calculated by the following formula:

$$Ei = - \sum_{j \in L} presision(i,j)$$
$$* \log\big(precision(i,j)\big)$$

The total entropy for a set of cluster is calculated as the sum of entropies for each cluster weighted by the size of each cluster:

$$Entropy_C = \sum_{i \in C} \left( \left(\frac{Ni}{N}\right) * Ei \right)$$

We need to maximize the purity measure and minimize the entropy of clusters in order to accomplish high quality clustering results.

## 5. RESULTS AND CONCLUSION

In this paper we present a new approach to document clustering based on topic maps representation of the documents. The inferred knowledge from the topic maps representation is used to define the similarity measure between a pair of documents. This measure is used to cluster the set of documents by using hierarchical agglomerative clustering (HAC). The experimental results show that the TMHC performs better than comparative algorithms of this study in term of quality of the clusters produced. An increased in cluster purity clearly established the fact that topic maps inherently capture the semantics of the documents. The experimental setup of the four well-know reported algorithms are implemented and tested on the datasets defined for this evaluation the F-measures for the algorithms are reported in Table 2, along with a graphical pattern of the measure. The proposed approach clearly had shown an improvement in all test cases. The F-Measure for dataset D4, which is a subset of NEWS20 produced a significant improvement. A further analysis of this reveals that when the document contains multi-topics like in D4, The suggested approach is exceptionally good.

**Table 2: F-Measure CFSW, FIHC, BKM and TMCH**

|  | CFSW | FIHC | BKM | TMHC |
|---|---|---|---|---|
| D1 | 0.57 | 0.63 | 0.66 | **0.68** |
| D2 | 0.61 | 0.65 | 0.63 | **0.68** |
| D3 | 0.68 | 0.88 | 0.76 | **0.89** |
| D4 | 0.71 | 0.74 | 0.74 | **0.78** |
| D5 | 0.69 | 0.69 | 0.66 | **0.72** |
| D6 | 0.77 | 0.76 | 0.72 | **0.78** |

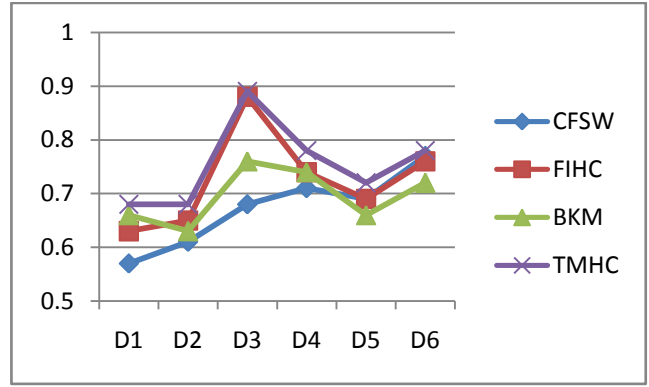

**Figure 2: F-Measure for CFSW, FIHC, BKM and TMHC**

The next measure is the purity of the clustering algorithms; the proposed algorithm had evidently produced better quality of the cluster. The improvement is significant in Reuters 21578 due to the fact that it contains large number of classes in the experimental data. This shows that the proposed algorithm is effective in clustering datasets with large unknown classes.

**Table 3: Purity for CFSW, FIHC, BKM and TMCH**

|  | CFSW | FIHC | BKM | TMHC |
|---|---|---|---|---|
| D1 | 0.71 | 0.76 | 0.65 | **0.8** |
| D2 | 0.72 | 0.78 | 0.64 | **0.81** |
| D3 | 0.72 | 0.78 | 0.66 | **0.81** |
| D4 | 0.69 | 0.7 | 0.68 | **0.75** |
| D5 | 0.74 | 0.76 | 0.7 | **0.77** |
| D6 | 0.77 | 0.81 | 0.73 | **0.84** |

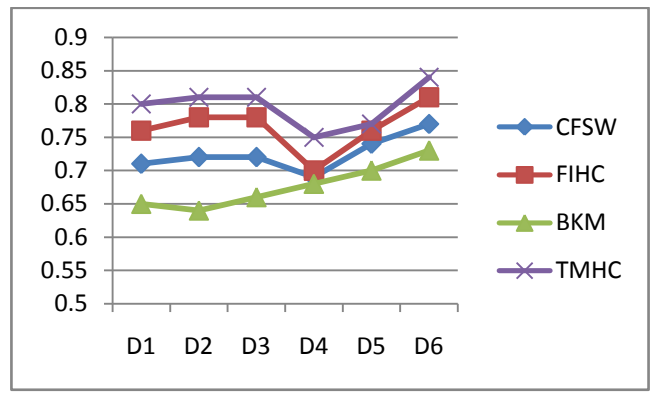

**Figure 3: Purity for CFSW, FIHC, BKM and TMHC**

The Entropy is another important measure for document clustering. Table 4 along with Figure 4 exhibits the results produced during the experiments. A clear reduction in entropy established the fact that the approach produces better clusters quality.

**Table 4: Entropy for CFSW, FIHC, BKM and TMCH**

|      | CFSW | FIHC | BKM  | TMHC   |
|------|------|------|------|--------|
| D1   | 0.22 | 0.21 | 0.27 | **0.2**  |
| D2   | 0.31 | 0.29 | 0.29 | **0.29** |
| D3   | 0.31 | 0.3  | 0.31 | **0.3**  |
| D4   | 0.26 | 0.21 | 0.26 | **0.2**  |
| D5   | 0.22 | 0.17 | 0.23 | **0.17** |
| D6   | 0.24 | 0.19 | 0.22 | **0.18** |

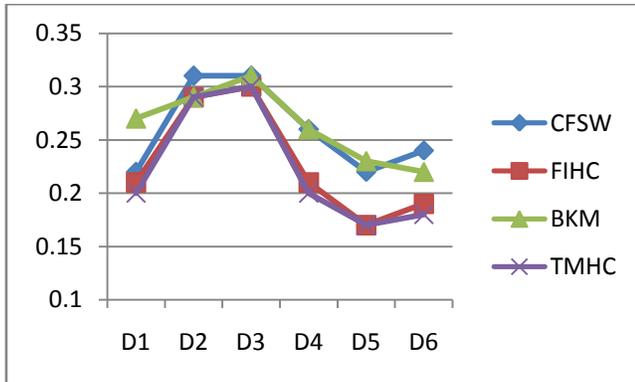

**Figure 4: Entropy for CFSW, FIHC, BKM and TMHC**

# 6. ACKNOWLEDGEMENT

We would like to acknowledge the anonyms reviewers for their valuable comments on the submitted paper. We would also like to thanks National University of Computer & Emerging Sciences, Islamabad for their kind support in this project. The Computer Science Department of Karachi Campus, along with The HoD provides every possible support to implement this project.

# 7. REFERENCES

[1] Jain, A. K., Murty, M. N., and Flynn, P. J., "Data Clustering: a review," ACM Computing Survey, pp. 264-323, 1999.

[2] Campi, A. and Ronchi, S., "The Role of Clustering in Search Computing ," in 20th International Workshop on Databases and Expert Systems Application , Linz, Austria, 2009, pp. 432-436.

[3] Cutting, D. R., Karger, D. R., Pedersen, J. O., and Tukey, J. W., "Scatter/Gather: A Cluster-based Approach to Browsing Large Document Collections," in Fifteenth Annual International ACM SIGIR Conference, June 1992, pp. 318-329.

[4] Hearst, M. A. and Pedersen, J. O., "Reexamining the cluster hypothesis: scatter/gather on retrieval results," in 19th annual international ACM SIGIR conference on Research and development in information retrieval, Zurich, Switzerland , 1996, pp. 74-84.

[5] Hammouda, K.M. and Kamel, M.S. , "Efficient Phrase-Based Document Indexing for Web Document Clustering," IEEE Transaction on Knowledge and Data Engineering, vol. 16, no. 10, pp. 1279-1296, 2004.

[6] Hung, C. and Xiaotie, D., "Efficient Phrase-Based Document Similarity for Clustering," IEEE Transaction on Knowledge and Data Engineering, vol. 20, no. September, pp. 1217-1229, 2008.

[7] Fung, B.C.M., Wang, K., and Ester, M., "Hierarchical document clustering using frequent Itemsets," Proceedings of SIAM International Conference on Data Mining, 2003.

[8] Soon, M. C. , John, D. H., and Yanjun, L., "Text document clustering based on frequent word meaning sequences," Data & Knowledge Engineering, vol. 64, pp. 381-404, 2006.

[9] Pepper, S., "Topic Maps," Encyclopedia of Library and Information Sciences, Third Edition 2010

[10] Pepper, S.; Moore, G., Eds. XML Topic Maps (XTM) 1.0; TopicMaps.Org 2001, http://www.topicmaps.org/xtm/1.0/

[11] Maicher, L.; Garshol, L.M.; Eds. Scaling Topic Maps; In Third International Conference on Topic Maps Research and Applications, TMRA 2007, Leipzig, Germany, October 2007; Springer-Verlag: Berlin, Heidelberg

[12] Steianbach, M., Karypis, G. , and Kumar, V., "A comparison of document clustering techniques," in KDD-Workshop on Text Mining , 2000.

[13] Rafi, M.,Maujood, M. ,Fazal, M. M., Ali, .M.,"Acomparison of two suffix tree-based document clustering algorithms", The Second IEEE ICIET,14-16 June Karachi, Pakistan, 2010.

[14] Document on Wandora Implementation and Usage. Can be found at http://www.wandora.org/wandora/wiki/index.php?title=Documentation